\documentclass[aps,prehardcore,reprint,showpacs,superscriptaddress]{revtex4-1}
\usepackage{amsmath,amssymb,physics}
\usepackage{graphicx}
\usepackage{dsfont}
\usepackage{bm}
\usepackage{xcolor}

\usepackage{epstopdf}

\usepackage[colorlinks=true,linkcolor=blue,citecolor=blue,urlcolor=blue]{hyperref}
\usepackage[all]{hypcap} 

\newcommand{\<}{\left\langle}
\renewcommand{\>}{\right\rangle}
\renewcommand{\(}{\left(}
\renewcommand{\)}{\right)}
\renewcommand{\[}{\left[}
\renewcommand{\]}{\right]}

\newcommand{\kB}{k_{\rm\scriptscriptstyle B}}
\newcommand{\rms}{\rm\scriptscriptstyle}
\newcommand{\rhoL}{\rho_{\rm\scriptscriptstyle L}}
\newcommand{\rhoR}{\rho_{\rm\scriptscriptstyle R}}
\newcommand{\rhoB}{\rho_{\rm\scriptscriptstyle b}}

\DeclareMathOperator*{\argmin}{argmin}
\DeclareMathOperator*{\argmax}{argmax}
\newcommand{\vw}{v_{\rm w}}
\newcommand{\jss}{j_{\rm ss}}
\newcommand{\varrhoss}{\varrho_{\rm ss}}
\newcommand{\varrhoeq}{\varrho_{\rm eq}}
\newcommand{\lw}{\lambda_{\rm w}}
\newcommand{\lf}{\lambda_{\rm f}}
\newcommand{\lb}{\lambda_{\rm b}}
\newcommand{\umf}{U_{\rm mf}}
\newcommand{\uint}{u^{\rm\scriptscriptstyle int}}
\newcommand{\uintY}{u^{\rm\scriptscriptstyle int}_{\rm\scriptscriptstyle Y}}
\newcommand{\AY}{A_{\rm\scriptscriptstyle Y}}
\newcommand{\Uint}{U^{\rm int}}
\newcommand{\fext}{f^{\rm ext}}
\newcommand{\fint}{f^{\rm int}}
\newcommand{\fintss}{f^{\rm int}_{\rm ss}}

\begin{document} 

\title[]{Nonequilibrium transport and phase transitions\\ in driven diffusion of interacting particles}

\author{Dominik Lips}
\email[]{dlips@uos.de}
\affiliation{Universit{\" a}t Osnabr{\" u}ck, Fachbereich Physik,
  Barbarastra{\ss}e 7, D-49076 Osnabr{\" u}ck, Germany}

\author{Artem Ryabov}
\email[]{rjabov.a@gmail.com}
\affiliation{Charles University, Faculty of Mathematics and Physics,
  Department of Macromolecular Physics, V Hole\v{s}ovi\v{c}k\'ach 2,
  CZ-18000 Praha 8, Czech Republic}

\author{Philipp Maass}
\email[]{maass@uos.de}
\affiliation{Universit{\" a}t Osnabr{\" u}ck, Fachbereich Physik,
  Barbarastra{\ss}e 7, D-49076 Osnabr{\" u}ck, Germany}

\date{January 24, 2020; revised February 13, 2020} 

\begin{abstract}
Driven diffusive systems constitute paradigmatic models of
nonequilibrium physics. Among them, a driven lattice gas known as the
asymmetric simple exclusion process (ASEP) is the most prominent
example for which many intriguing exact results have been
obtained. After summarizing key findings, including the mapping of the
ASEP to quantum spin chains, we discuss the recently introduced Brownian
asymmetric simple exclusion process (BASEP) as a related class of
driven diffusive system with continuous space dynamics.  In the BASEP,
driven Brownian motion of hardcore-interacting particles through
one-dimensional periodic potentials is considered. We study whether
current-density relations of the BASEP can be considered as generic
for arbitrary periodic potentials and whether repulsive particle
interactions other than hardcore lead to similar results. Our findings
suggest that shapes of current-density relations are generic for
single-well periodic potentials and can always be attributed to the
interplay of a barrier reduction, blocking and exchange symmetry
effect.  This implies that in general up to five different phases of
nonequilibrium steady states are possible for such potentials.  The
phases can occur in systems coupled to particle reservoirs, where the
bulk density is the order parameter.  For multiple-well periodic
potentials, more complex current-density relations are possible and
more phases can appear.  Taking a repulsive Yukawa potential as an
example, we show that the effects of barrier reduction and blocking on
the current are also present.  The exchange symmetry effect requires
hardcore interactions and we demonstrate that it can still be
identified when hardcore interactions are combined with weak Yukawa
interactions.  The robustness of the collective dynamics in the BASEP
with respect to variations of model details can be a key feature for a
successful observation of the predicted current-density relations in
actual physical systems.

\end{abstract}

\maketitle  

\section{Introduction}
\label{sec:introduction}

Driven diffusive systems of interacting particles constitute an
important class of systems to study fundamental aspects of
nonequilibrium physics. This holds in particular for one-dimensional
models, where exact analytical derivations are possible or reliable
approximations are known, for example, when information about exact
equilibrium properties can be utilized for the treatment of
nonequilibrium states.

A prominent model in the field of driven diffusive systems is the
asymmetric simple exclusion process (ASEP), where particles hop
between nearest-neighbor sites of a lattice with a bias in one
direction and where the sole interaction between particles is a mutual
site exclusion, implying that a lattice site cannot be occupied by
more than one particle \cite{Derrida:1998, Schuetz:2001}. In the ASEP
on a one-dimensional lattice with $L$ sites and periodic boundary
conditions, i.e.\ a ring of $L$ sites, particles jump to vacant
nearest-neighbor sites with rates $\Gamma_+$ and $\Gamma_-$ in
clockwise and counterclockwise direction, respectively, where
$\Gamma_+>\Gamma_-$ for a bias in clockwise direction.  In a
corresponding open system with $L$ sites, where the leftmost and
rightmost lattice site can exchange particles with reservoirs L and R,
respectively, additional rates $\Gamma^{\rm\scriptscriptstyle L}_{\rm
  in}$, $\Gamma^{\rm\scriptscriptstyle R}_{\rm in}$ and
$\Gamma^{\rm\scriptscriptstyle L}_{\rm out}$,
$\Gamma^{\rm\scriptscriptstyle R}_{\rm out}$ specify the corresponding
rates for particle injection and ejection.  Many properties of the
ASEP can be inferred from the even simpler totally asymmetric simple
exclusion process (TASEP) with unidirectional transport
($\Gamma_-=0$).

Stochastic processes in driven lattice gases are described by a master
equation for the probabilities of particle configurations, which can
be viewed also as the occupation number representation of a
Schr\"odinger equation in imaginary time \cite{Gwa/Spohn:1992,
  Sandow/Trimper:1993}.  This leads to some interesting connections to
quantum systems with in general non-Hermitian Hamilton operator $H$
\cite{Gwa/Spohn:1992, Sandow:1994, Henkel/Schuetz:1994,
  Schuetz:2001}. As an example, we recapitulate in the Appendix the
connection of the ASEP with periodic boundary conditions to the $XXZ$
quantum spin chain with non-Hermitian boundary conditions
\cite{Henkel/Schuetz:1994}.  Spin chains are often used to study
fundamental aspects of nonequilibrium quantum physics. Several
examples related to current problems, in particular to questions of
equilibration in non-integrable spin chain models, can be found in
this special issue.

The ASEP has been intensively studied in the past. Let us summarize
here some of the most important findings for the ASEP and variants of
it:\\[-2ex]
\begin{list}{--}{\setlength{\leftmargin}{1em}\setlength{\itemsep}{0.5ex}\setlength{\parsep}{0cm}
\setlength{\topsep}{0ex}}

\item Using the Bethe ansatz for corresponding quantum spin chain
  models, or a construction in terms of matrix product states, exact
  results for microstate distributions in nonequilibrium steady states
  (NESS) could be derived \cite{Derrida:1998, Schuetz:2001,
    Blythe/Evans:2007}. Matrix product states in principle exist for
  driven lattice gases with arbitrary nearest-neighbor interactions
  \cite{Krebs/Sandow:1997}, although their explicit construction may
  be difficult.

\item Based on the exact approaches for deriving distribution of
  microstates in NESS, large deviation functions for fluctuations of
  time-averaged densities and currents were derived
  \cite{Derrida/Lebowitz:1998, Derrida/etal:2002,
    Krapivsky/etal:2014}. They have been computed also for
  coarse-grained descriptions by the macroscopic fluctuation theory
  \cite{Bertini/etal:2015}. Large deviation functions are argued to
  play a similar role for time-averaged quantities in NESS as the free
  energy in equilibrium systems \cite{Touchette:2009}.  They can
  exhibit singularities \cite{Bertini/etal:2005,
    Bodineau/Derrida:2005, Appert-Rolland/etal:2008, Baek/etal:2017},
  sometimes referred to as ``dynamical phase transitions'', which for
  certain classes of systems are caused by a violation of an
  ``additivity principle'' \cite{Bodineau/Derrida:2004}.

\item The Bethe ansatz turned out to be a valuable tool also for
  deriving microstate distributions of non-steady states
  \cite{Schuetz:1997, Tracy/Widom:2008}.  The propagator for the
  microstate time evolution in the ASEP was related to integrated
  Fredholm determinants \cite{Tracy/Widom:2008b} and led to the
  derivation of the Tracy-Widom distribution of random matrix theory
  for the asymptotic behavior in case of a step initial condition
  \cite{Tracy/Widom:2009}.  This result generalized an earlier one
  derived for the TASEP \cite{Johansson:2000} and proved that the
  propagation of density fluctuations in the ASEP belongs to the
  Kardar-Parisi-Zhang (KPZ) universality class
  \cite{Kardar/etal:1986}.

\item In open systems coupled to particle reservoirs, phase
  transitions between NESS occur \cite{Krug:1991, Schuetz/Domany:1993,
    Kolomeisky/etal:1998, Brzank/Schuetz:2007}, where, upon change of
  control parameters characterizing the system-reservoir couplings,
  the bulk density $\rhoB$ changes discontinuously, or its derivative
  with respect to the control parameters.  Knowing the density
  dependence of the steady-state bulk current $\jss(\rho)$, e.g.\ from
  results for a system with periodic boundary conditions, all possible
  NESS phases with bulk density $\rhoB$ are predicted by the extremal
  current principles \cite{Krug:1991, Kolomeisky/etal:1998,
    Popkov/Schuetz:1999}
\begin{equation}
\rhoB=\left\{\begin{array}{l@{\hspace{1em}}l}
\displaystyle\argmin_{\rho_-\le\rho\le\rho_+}\{\jss(\rho)\}\,,
& \rho_-\le\rho_+\,,\\[3ex]
\displaystyle\argmax_{\rho_+\le\rho\le\rho_-}\{\jss(\rho)\}\,,
& \rho_+\le\rho_-\,.
\end{array}\right.
\label{eq:extremal-current-principles}
\end{equation}
Here $\rho_-$ and $\rho_+$ can be any densities bounding a
monotonically varying region encompassing the plateau part with bulk
density $\rhoB$ (which may strictly exist only in the thermodynamic
limit of infinite system size).  Which of the phases predicted by
Eq.~\eqref{eq:extremal-current-principles} really occurs for a given
control scheme of system-reservoir couplings, is given by the
dependence of $\rho_-$ and $\rho_+$ on respective control parameters.

The extremal current principles can be reasoned based on the
consideration of shock front motions \cite{Kolomeisky/etal:1998,
  Popkov/Schuetz:1999, Hager/etal:2001}, or by resorting to a
decomposition of the steady-state current into its drift and diffusive
part inside the region of monotonically varying density profile
\cite{Krug:1991}.  Because these reasonings do not require specific
properties of the ASEP, they are quite generally valid for driven
diffusive systems coupled to particle reservoirs.  This includes
driven lattice gases with interactions other than site exclusion
\cite{Antal/Schuetz:2000, Hager/etal:2001, Dierl/etal:2012,
  Dierl/etal:2013}, systems with continuous space-dynamics and systems
with periodic space structure and/or time-periodic driving, when
considering period-averaged densities \cite{Dierl/etal:2014}.  For
specific system-reservoir couplings termed ``bulk-adapted'' it is
possible to parameterize the exchange of particles by reservoir
densities such that all possible NESS phases must appear. The
bulk-adapted couplings can be determined by a general method for
driven lattice gases with short-range interactions
\cite{Dierl/etal:2013, Dierl/etal:2014}.

\item For random and non-Poissonian hopping rates, Bose-Einstein type
  condensations of vacancies can occur in front of the slowest
  particle with smallest jump rate \cite{Evans:1996,
    Concannon/Blythe:2014}.

\item Coarse-grained continuum descriptions of the ASEP and of
  multilane variants \cite{Popkov/Salerno:2004} give rise to an
  infinite discrete family of nonequilibrium universality classes in
  nonlinear hydrodynamics, where density fluctuations spread in time
  by power laws with exponents given by the Kepler ratios of
  consecutive Fibonacci numbers \cite{Popkov/etal:2015}.  This
  includes the KPZ class, for which an exact expression for the
  scaling function was derived
  \cite{Praehofer/Spohn:2004}. Predictions of the theory of
  fluctuating nonlinear hydrodynamics were recently confirmed in an
  exact treatment by considering a two-species exclusion process
  \cite{Chen/etal:2018}.

\end{list}

As for applications, the ASEP appears as a basic building block in
manifold descriptions of biological traffic
\cite{Schadschneider/etal:2010, Chou/etal:2011}. In fact, the ASEP was
introduced first to describe protein synthesis by ribosomes
\cite{MacDonald/etal:1968}, and it is frequently used in connection
with the motion of motor proteins along microtubules or actin tracks
\cite{Kolomeisky:2013, Appert-Rolland/etal:2015}.  An in-vitro study
with fluorescently labeled single-headed kinesin motors moving along a
microtubule provided experimental evidence for a state of coexisting
phases with different motor densities
\cite{Nishinari/etal:2005}. Other applications concern vehicular
traffic \cite{Appert-Rolland/etal:2011, Foulaadvand/Maass:2016},
diffusion of ions through cell membranes \cite{Hille:2001} and of
molecules through nanopores \cite{Cheng/Bowers:2007,
  Dvoyashkin/etal:2014}, and electron transport along molecular wires
in the incoherent classical limit \cite{Nitzan:2001,
  Berlin/etal:2001}.  However, a direct experimental realization of
the ASEP is difficult, because of its discrete nature.  Hence, it is
important to see whether the nonequilibrium physics in the ASEP is
reflected in models with continuous space dynamics.

For a single particle, it is well known that effective hopping
transport emerges from an overdamped Brownian motion in a periodic
potential with amplitude much larger than the thermal energy.  The
particle can be viewed to jump between neighboring wells on a
coarse-grained time scale with a rate determined by the inverse
Kramers time \cite{Risken:1985}. One is thus led to ask whether the
driven diffusion of many hardcore interacting particles in a periodic
energy landscape can reflect the driven lattice gas dynamics in the
ASEP. To answer this question, we recently introduced a corresponding
class of nonequilibrium processes termed Brownian ASEP (BASEP)
\cite{Lips/etal:2018, Lips/etal:2019}, where hard spheres with
diameter $\sigma$ are driven through a periodic potential with
wavelength $\lambda$ by a constant drag force $f$. For a sinusoidal
external potential, we found that the current-density relation of the
ASEP is indeed recaptured in the BASEP, but only for a limited range
of particle diameters $\sigma$.  For other $\sigma$, quite different
behaviors are obtained.

The nonequilibrium physics of the BASEP should be explorable directly
by experiment, for example in setups utilizing advanced techniques of
microfluidics and optical and/or magnetic micromanipulation
\cite{Arzola/etal:2017, Skaug/etal:2018, Schwemmer/etal:2018,
  Stoop/etal:2019, Misiunas/Keyser:2019}. This includes arrangements
where the particles are driven by traveling-wave potentials
\cite{Straube/Tierno:2013}. Many of the new collective transport
properties seen in the BASEP can be even identified by studying local
dynamics of individual transitions between potential wells
\cite{Ryabov/etal:2019}.

In this work we address the question how the current-density relations
found for the BASEP in a sinusoidal external potential are affected
when considering different external potentials and short-range
interactions other than hardcore exclusions.  Our investigation for
the different external potentials are carried out based on the
small-driving approximation introduced in Refs.~\cite{Lips/etal:2018}
and \cite{Lips/etal:2019}.  With respect to short-range interactions
other than hardcore exclusions, we focus on a Yukawa pair potential.
It is shown that the current-density relation for single-well periodic
potentials and for the Yukawa interaction has similar features as that
of the BASEP.  This suggests that the BASEP can serve as a reference
model for a wide class of external periodic potentials and pair
interactions.

In addition we extend a former analysis to prove that current
reversals cannot occur in systems driven by a constant drag and by
traveling waves. These proofs are based on an exact calculation of the
total entropy production in corresponding NESS for particles with
arbitrary pair interactions.  Current reversals refer to steady
states, where particle flow is opposite to the external bias.  They
were reported for lattice models \cite{Jain/etal:2007, Slanina:2009a,
  Slanina:2009b, Chaudhuri/Dhar:2011, Dierl/etal:2014} and were
recently found experimentally in a rocking Brownian motor
\cite{Schwemmer/etal:2018}. Their absence in traveling-wave driven
systems was conjectured based on simulation results and a perturbative
expansion of the single-particle density in the NESS around its
period-averaged value \cite{Chaudhuri/etal:2015}.

The paper is organized as follows. In Sec.~\ref{sec:current-density}
we present an analytical treatment of densities and currents for the
overdamped one-dimensional Brownian motion of particles with arbitrary
pair interactions. This section partly summarizes results presented
earlier \cite{Lips/etal:2018, Lips/etal:2019} and introduces the
small-driving approximation used subsequently for our investigation of
hardcore interacting particles. It also contains our proofs on the
absence of current reversals for general pair interactions.  In
Sec.~\ref{sec:reference-BASEP} we outline our findings for the BASEP
with sinusoidal external potential, and in
Sec.~\ref{sec:impact-external-potential} we contrast them with results
for a Kronig-Penney and triple-well periodic potential.  In
Sec.~\ref{sec:impact-interactions} we discuss our results for the
Yukawa interaction. Section~\ref{sec:conclusions} concludes the paper
with a summary and outlook.

\section{Current-density relations: Analytical results}
\label{sec:current-density}
The overdamped single-file Brownian motion of $N$ particles in a
periodic potential $U(x)=U(x+\lambda)$ with pair interaction under a
constant drag force $f$ is described by the Langevin equations
\begin{align}
\frac{\dd x_i}{\dd t} = 
\mu \left(f+\fint_i-
\frac{\partial U(x_i)}{\partial x_i}\right) + \sqrt{2D} \, \eta_i(t) \,,
\label{eq:langevin}
\end{align}
where $\mu$ and $D=\mu \kB T$ are the bare mobility and diffusion
coefficient, $\kB T$ is the thermal energy, and $\fint_i$ is the
interaction force on the $i$th particle. The $\eta_i(t)$ are
independent and $\delta$-correlated Gaussian white noise processes
with zero mean and unit variance, $\langle \eta_i \rangle =0$ and
$\langle \eta_i(t) \eta_j(t') \rangle = \delta_{ij}
\delta(t-t')$. Unless noted otherwise, we consider closed systems with
periodic boundary conditions, which means that the particles are
dragged along a ring.

Hardcore interactions imply the boundary conditions $|x_{i} - x_{j}|
\geq \sigma$, i.e.\ overlaps between neighboring particles are
forbidden. For the BASEP with only hardcore interactions, these
boundary conditions must be taken into account, while the interaction
force $f_i^{\rm \scriptscriptstyle int}$ can be set to zero in
Eq.~\eqref{eq:langevin}.  We define the density as a (dimensionless)
filling factor of the potential wells, i.e.\ by $\rho=N/M$, where $M$
denotes the total number of periods of $U(x)$.  The system length is
$L=M\lambda$ and the number density is $\rho/\lambda$. For hardcore
interacting particles of size $\sigma$, the filling factor $\rho$ has
the upper bound $1/\sigma$.
  
The joint probability function (PDF) of the particle center
coordinates $\boldsymbol{x}=(x_1, \ldots , x_N )$ evolves in time
according to the $N$-particle Smoluchowski equation \footnote{To
  implement the periodic boundary conditions, we assume an ordered
  initial configuration $0 \leq x_{1} \leq x_{2} \ldots \leq x_N < L$,
  and introduce two fictive particles with enslaved coordinates
  $x_0=x_N-L$ and $x_{N+1} = x_1+L$, which implies $x_N - x_1 <
  L-\sigma$.},
\begin{equation}
\label{eq:FokkerPlanck}
\frac{\partial p_N(\boldsymbol{x},t ) }{\partial t} = - \div
\boldsymbol{J}(\boldsymbol{x},t )\,,
\end{equation}
where the divergence operator acts on the probability current vector
$\boldsymbol{J}(\boldsymbol{x},t )$ with the $i$th component given by
\begin{align} 
\label{eq:Ji}
\begin{split} 
J_i(\boldsymbol{x},t ) =\, &  
\mu \[ f - \frac{\partial U(x_i)}{\partial x_i} + 
\fint_i(\boldsymbol{x}) \] p_N(\boldsymbol{x},t ) \\  
& - D \frac{\partial p_N(\boldsymbol{x},t )}{\partial x_i}\,. 
\end{split} 
\end{align} 
The first term describes the drift probability current caused by all
forces acting on the $i$th particle and the second term gives the
diffusive current.  The interaction force is assumed to be
conservative and due to pair interactions $\uint(x_i,x_j)$,
i.e.\ $\fint_i(\boldsymbol{x})=-\partial\Uint(\boldsymbol{x})/\partial
x_i$ with
\begin{equation} 
\label{eq:Uint}
\Uint(\boldsymbol{x}) = \frac{1}{2}\sum_{i\ne j}^N \uint(x_i , x_j )\,.
\end{equation} 

Additional hardcore interactions are not included in the
potential~\eqref{eq:Uint} but are incorporated into the dynamics by
requiring no-flux (reflecting) boundary conditions
\begin{equation} 
\label{eq:hardcoreBC}
\left. \[ J_{i}(\boldsymbol{x},t) - J_{i+1}(\boldsymbol{x},t) \] \right|_{x_{i+1}=x_i+\sigma}=0, 
\end{equation} 
if neighboring particles hit each other.  These boundary conditions
ensure conservation of an initial ordering $x_1<x_2<\ldots<x_N$ of the
particle positions for all times.

\subsection{Exact current-density relation}
\label{subsec:small-driving}
The local density is
\begin{equation}
\varrho (x,t) = \< \sum_{i=1}^{N} \delta[x-x_i(t)] \> , 
\end{equation}
where the average is taken with respect to the solution of the
Smoluchowski equation~\eqref{eq:FokkerPlanck} subject to some initial
condition. It satisfies the continuity equation
\begin{equation} 
\label{eq:continuityEQ}
\frac{\partial \varrho (x,t)}{\partial t} = - \frac{\partial j(x,t)}{\partial x} , 
\end{equation}
with the particle current density given by \cite{Lips/etal:2018}  
\begin{equation}
\label{eq:jxt}
j(x,t) = \mu \[f^{\rm ext}(x) + \fint(x,t)  \] \varrho(x,t)  
- D \frac{\partial \varrho(x,t)}{\partial x}\,.
\end{equation} 
Here we introduced the total external force
\begin{equation}
\fext(x)= f - \frac{\partial U(x)}{\partial x}\,. 
\end{equation}
The local interaction force $\fint(x,t)$ in Eq.~\eqref{eq:jxt} is given by 
\begin{equation} 
\label{eq:fint}
\fint(x,t) = \frac{1}{\rho(x,t)} \int_0^L \dd y\, f_2(x,y)\rho_2(x,y,t),
\end{equation}
where  
\begin{equation}
\label{eq:rho2}
\varrho_2(x,y,t) = \< \sum_{i\ne j}^{N}\delta[x-x_i(t)]\, \delta[y-x_j(t)] \> ,
\end{equation} 
is the two-point local density, and $f_2(x,y)$ is the interaction
force of a particle at position $y$ on a particle at position $x$. It
can by expressed as a sum of two distinct contributions:
\begin{align}
\label{eq:f2}
f_2(x,y) =\, & \kB T \[ \delta(y-x+\sigma) - \delta(x-y-\sigma) \]
 - \frac{\partial\uint(x,y)}{\partial x} .
\end{align}
The first term is due to a positive and a negative force, if
a particle is in contact with other particles at
positions $x-\sigma$ and $x+\sigma$, respectively.
We note that the $\delta$-functions
should not be interpreted as a derivative of a rectangular potential
barrier of height $\kB T$. Instead, they are a consequence of the
noncrossing boundary conditions~\eqref{eq:hardcoreBC}
\cite{Lips/etal:2018}. 
The amplitude in front of the $\delta$-functions
must be an energy on dimensional reasons, for which $\kB T$ is the
only relevant scale. It corresponds to the typical collision energy 
due to the thermal noise. The
second term in Eq.~\eqref{eq:f2} is the force due to the interaction
potential~\eqref{eq:Uint}. 

In the steady state of a closed system with periodic boundary
conditions, the density profile is time-independent and periodic,
$\varrhoss(x+\lambda )=\varrhoss(x)$ and the current constant
everywhere in the system. It follows directly from Eq.~\eqref{eq:jxt}
\cite{Lips/etal:2018}
\begin{equation}
\label{eq:jst}
\jss(\rho,\sigma ) =  
\frac{\mu \[ f+\frac{1}{\lambda}\int_0^\lambda \dd x\, \fint_{\rm ss}(x) \]}
{\frac{1}{\lambda}\int_0^\lambda \dd x\, \varrhoss^{-1}(x)}\,.
\end{equation}

Up to this point no approximation has been made. The exact value of
the steady-state current~\eqref{eq:jst} depends on both $\varrhoss(x)$
and the steady-state limit of the two-point
density~\eqref{eq:rho2}. However, derivation of the two densities in
NESS represents a challenging problem, which can be solved in a few
special cases only. Therefore, to proceed further, we need to develop
an appropriate approximate theory.

\subsection{Small-driving approximation}
\label{subsec:small-driving}
For hardcore interactions, the small-driving approximation (SDA)
turned out to be particularly successful in capturing qualitative
behaviors of $\jss(\rho,\sigma)$ \cite{Lips/etal:2019}. 
The approximation is carried out in two steps. Firstly, we linearize the
current~\eqref{eq:jst} with respect to $f$,
\begin{equation}
\label{eq:jst_linear}
\jss(\rho,\sigma) \sim  
\frac{ \( 1+ \chi  \) \mu f}{\frac{1}{\lambda}\int_0^\lambda \dd x\, \varrhoeq^{-1}(x)}, 
\quad f \to 0, 
\end{equation}
where the response coefficient reads
\begin{equation} 
\chi = \left. \frac{\partial }{\partial f} \[ \frac{1}{\lambda}\int_0^\lambda \dd x\, \fintss(x)  \] \right|_{f=0},
\end{equation}  
and, secondly, we approximate the linear-response
expression~\eqref{eq:jst_linear} by setting $\chi=0$.

The ad-hoc $\chi=0$ approximation works well in an extended region of
particle sizes except for a narrow range $\sigma \approx \lambda/2$
\cite{Lips/etal:2019}. The equilibrium density profile is obtained by
minimizing the exact density functional for hard rods
\cite{Percus:1976},
\begin{align}
\Omega[\varrho(x)] = &\int\limits_0^\lambda \dd x\, \varrho(x) \biggl\{
U(x) - \mu_{\rms ch} \nonumber \\
&{}- \kB T \left[ 1 - \ln\left(
\frac{\varrho(x)}{1 - \eta(x)} \right) \right]\biggr\}, 
\label{eq:percus}
\end{align} 
where $\mu_{\rms ch}$ is the chemical potential, and
\begin{equation}
\eta(x) = \int\limits_{x-\sigma}^x \dd y\, \varrho(y) \, .
\label{eq:eta}
\end{equation}
The minimization yields the structure equation
\begin{align}
0=\frac{\delta \Omega[\varrho]}{\delta \varrho}
\bigg|_{\varrho=\varrho_{\rms eq}} &= \kB T\ln\left[
\frac{\varrho_{\rms eq}(x)}{1-\eta_{\rms
    eq}(x)}\right] \label{eq:structure-equation}\\ 
&\hspace{-2em}+\kB T\int\limits_{x}^{x+\sigma} \dd y \frac{\varrho_{\rms
    eq}(y)}{1-\eta_{\rms eq}(y)}
+ [U(x) - \mu_{\rms ch}] \,,\nonumber
\end{align}
which we discretized and solved numerically under periodic boundary
conditions [$\varrho_{\rms eq}(x)=\varrho_{\rm eq}(x+\lambda)$].  The
chemical potential $\mu_{\rms ch}$ was adjusted to give the desired
global density (filling factor) $\rho=\int_0^\lambda\dd x\,
\varrho_{\rms eq}(x)$.

\subsection{Entropy production and absence of current reversals}
\label{subsec:current-reversals}
Theory for hardcore interacting particles with
$\Uint(\boldsymbol{x})=0$ was the subject of our previous works on the
BASEP \cite{Lips/etal:2018, Ryabov/etal:2019, Lips/etal:2019}. Here we
extend the analysis to nonzero $\Uint(\boldsymbol{x})$. We start with
considerations related to the total entropy production:
\begin{equation}
\dot{S}_{\rm tot}(t) = \dot{S}_{\rm sys}(t) + \dot{S}_{\rm med}(t)\,, 
\end{equation}
where $\dot{S}_{\rm sys}(t)$ and $\dot{S}_{\rm med}(t)$ are the
entropy production in the system and surrounding medium.

For calculating the time derivative of $S_{\rm sys}(t) = - \kB \int
d^N\!x\, p_N(\boldsymbol{x},t) \ln p_N(\boldsymbol{x},t)$ we can
replace the time derivative of the PDF by the divergence of the
current according to \eqref{eq:FokkerPlanck}. After integrating by
parts of each individual term of the divergence, we get
\begin{equation}
\frac{\dot{S}_{\rm sys}(t)}{\kB } = - 
\int_\Omega \dd^N\!x \,\boldsymbol{J}(\boldsymbol{x},t ) \cdot \grad   \ln p_N(\boldsymbol{x},t),
\end{equation}
where $\Omega$ is the space of all system microstates consistent with
the hardcore constraints.  As the next step, we replace $\partial \ln
p_N /\partial x_i $ via Eq.~\eqref{eq:Ji}, which gives us two terms
\begin{equation}
\frac{\dot{S}_{\rm sys}(t)}{\kB } = 
\int_\Omega\dd^N\!x\,
\frac{ |\boldsymbol{J}(\boldsymbol{x},t )|^2 }{D p_N(\boldsymbol{x},t )}
- \int_\Omega\dd^N\!x\,
\frac{\boldsymbol{J}(\boldsymbol{x},t )\cdot \boldsymbol{F}(\boldsymbol{x} )}{\kB T}\,.
\end{equation}
Here we have introduced the total force
$\boldsymbol{F}(\boldsymbol{x})$ with components
$F_i(\boldsymbol{x})=\fext(x_i)+\fint_i(\boldsymbol{x})$.  The first
term is always positive and equal to the total entropy production
\cite{Seifert:2012}.  The second term, proportional to the mean
dissipated power, is the entropy production in the medium.

In the steady state, the system entropy is constant, $\dot{S}_{\rm
  sys}(t)=0$, and the total entropy production equal to the entropy
produced in the surrounding medium:
\begin{align} 
0\le\dot{S}_{\rm tot}&= \frac{1}{T} \int_\Omega\dd^N\!x\,
\boldsymbol{J}(\boldsymbol{x})\cdot \boldsymbol{F}(\boldsymbol{x}) 
\label{eq:dotstot1}\\
&\hspace{-2em}=\frac{1}{T} \sum_{i=1}^N \int_\Omega\dd^N\!x\,
J_i(\boldsymbol{x}) \[ f-\frac{\partial U(x_i)}{\partial x_i}
- \frac{\partial \Uint(\boldsymbol{x})}{\partial x_i} \]. \nonumber\\
\nonumber
\end{align} 
Here, each single-particle term simplifies after introducing the
current density $ j_i(x) =
\int_\Omega\dd^N\!x\,J_i(\boldsymbol{x})\delta(x_i-x)$, of the $i$th
particle, and by using that $j_i(x)=\jss/N$ in the steady state,
\begin{equation}
\label{eq:Smi}
\int_\Omega\dd^N\!x\,
J_i(\boldsymbol{x}) \[ f-\frac{\partial U(x_i)}{\partial x_i} \] 
 = \frac{\jss}{N} L f\,. 
\end{equation}
The sum over all interaction forces in Eq.~\eqref{eq:dotstot1} yields,
after integration by parts,
\begin{equation}
- \sum_{i=1}^N  \int_\Omega\dd^N\!x\,
J_i(\boldsymbol{x}) \frac{\partial \Uint(\boldsymbol{x})}{\partial x_i} =
\int_\Omega\dd^N\!x\,\, \Uint(\boldsymbol{x})\, \div \boldsymbol{J}(\boldsymbol{x})\,.
\end{equation} 
Because the divergence of the current is zero in the steady state,
this term vanishes. From Eqs.~\eqref{eq:dotstot1} and \eqref{eq:Smi}
we thus obtain the total entropy production in the Onsager form
(current times thermodynamic force)
\begin{equation}  
\label{eq:StotBASEP}
\dot{S}_{\rm tot} =  \, \frac{\jss f}{T} L  \geq 0. 
\end{equation} 
It is extensive in the system size and the numerator equals the mean
heat dissipated at any point of the system in the steady state. As a
consequence of the inequality in~\eqref{eq:StotBASEP}, the steady
state current must have the same sign as the drag force $f$.

\subsection{Entropy production in traveling-wave driven systems and current bounds}
\label{subsec:traveling-wave}
A feasible way to verify BASEP current-density relations in a
laboratory is to consider an equivalent ring system with the
traveling-wave (TW) external periodic potential $U(x-v_w t)$ and $f=0$
\cite{Siler/etal:2008, Straube/Tierno:2013, DiLeonardo/etal:2007,
  Curran/etal:2010, Siler/etal:2012}. In such a TW system, the $i$th
compomnent of the probability current vector is
\begin{align} 
\label{eq:JiTW}
\begin{split} 
J_i^{\rm TW}\!(\boldsymbol{x},t ) =\, &  
\mu\! \[- \frac{\partial U(x_i-v_w t)}{\partial x_i}\! 
+\! \fint_i(\boldsymbol{x}) \]\! p_N^{\rm TW}\!(\boldsymbol{x},t ) \\  
& - D \frac{\partial p_N^{\rm TW}(\boldsymbol{x},t )}{\partial x_i}. 
\end{split} 
\end{align} 

Under a Galilean transformation
\begin{equation}
\label{eq:Galilei}
x_i(t) = x_i^{\rm TW}(t) - v_{w} t. 
\end{equation} 
the TW system maps to the corresponding BASEP with potential $U(x)$
and constant drag force
\begin{equation}
\label{eq:v_w}
f=-\frac{v_w}{\mu}\,,
\end{equation}
provided the pair interaction potential $\uint(x,y)$ is a function of the particle distance $(x-y)$ only.
Local densities and currents of the two corresponding systems are related by \cite{Lips/etal:2019}
\begin{align}
\label{eq:rhoTW}
& \varrho^{\rm TW}(x,t) = \varrho(x-v_w t,t) , \\ 
\label{eq:jTW}
& j^{\rm TW}(x,t) = v_w \varrho(x-v_w t,t) + j(x-v_w t,t)\,.
\end{align}

A remarkable aspect of this mapping that has not been discussed in our
previous work relates to the fundamental difference of dissipations
(their physical origins and their values) in the two pictures. In
fact, hopping events that contribute positively to the dissipation
(total entropy production) in one picture, cause a decrease of the
dissipation in the other.

In the BASEP, the dissipation equals the average work done by a
constant non-conservative force on all the particles, see Eq.~\eqref{eq:StotBASEP}. 
In the TW system, there is no non-conservative force. Instead, each particle is acted upon by the
time-dependent force $[-U'(x-v_wt)]$ and the sum of these actions over
all particles gives the total power input into the system.  In the
steady state, this power is dissipated into the ambient heat bath via
friction. Therefore, in the TW model, the total entropy production
averaged over one period $\tau=\vw/\lambda$ reads (the bar denoting
period-averaging in time)
\begin{equation}
\overline{\dot{S}_{\rm tot}^{\rm TW}}= 
- \frac{1}{T} \sum_{i=1}^N \frac{1}{\tau} \int_0^\tau\!\! \dd t  \< \frac{\partial U(x_i-v_w t)}{\partial t} \>.
\end{equation} 

After some algebra similar to that in
Sec.~\ref{subsec:current-reversals} one obtains \cite{Lips/etal:2019}
\begin{equation}
\label{eq:StotTW}
\overline{\dot{S}_{\rm tot}^{\rm TW}}= \frac{v_w \overline{\jss^{\rm TW}}}{\mu T} L  \geq 0 , 
\end{equation}
which means that the period-averaged stationary current
$\overline{j_{\rm st}^{\rm TW}}$ must have the same sign as
$v_w$. Hence there are no current reversals in a TW system.

Furthermore, we can relate the TW current in Eq.~\eqref{eq:StotTW} to
the corresponding BASEP by taking the period-averaged form of
Eq.~\eqref{eq:jTW} in the steady state:
\begin{equation}
\label{eq:StotTW_jst}
\overline{\dot{S}_{\rm tot}^{\rm TW}}= \frac{v_w^2\rho + v_w \jss}{\mu T} L \geq 0 .
\end{equation}
Here, the two terms in the numerator have clear physical meanings. The
first contains the expression $v_w^2/\mu$ equal to the dissipated
power by a particle moving at constant velocity $v_w$ in the fluid
characterized by the friction coefficient $1/\mu$.  This term gives
the maximal possible dissipation in the TW system corresponding to the
case with no jumps over potential barriers where the motion of each
particle is exactly phase-locked with the TW potential $U(x-v_w t)$.
The second term contains the current in the corresponding BASEP and is
negative because $v_w$ and $f$ have opposite signs. Recalling
Eq.~\eqref{eq:v_w}, we see that the second term equals exactly the
dissipation in the BASEP~\eqref{eq:StotBASEP} up to the minus sign. 
It tells us that the total TW dissipation is diminished by the difference of the average
number of jumps over potential barriers in and against bias direction.

Overall, the inequality in Eq.~\eqref{eq:StotTW_jst} implies
\begin{equation}
0 \leq \jss(\rho,\sigma) \leq \mu f \rho\,,
\end{equation}
for $f>0$, i.e.\ we obtain the upper bound $\mu f \rho$ for the
current, while the lower bound follows from Eq.~\eqref{eq:StotBASEP}
as already discussed.

\begin{figure}[b!]
\includegraphics[width=0.45\textwidth]{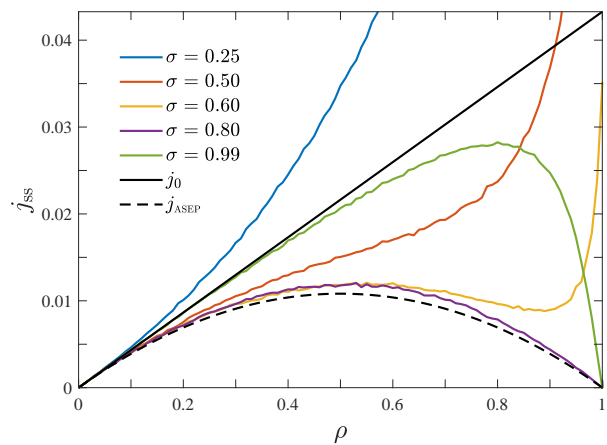}
\caption{Simulated steady-state current $\jss$ in the BASEP with
  cosine potential [Eq.~\eqref{eq:cosine-potential}] as function of
  the density $\rho$ for different particle sizes $\sigma$. The solid
  black line marks the current of non-interacting particles
  $j_0(\rho)=v_0\rho$, and the dashed line the current-density
  relation $j_{\rms ASEP}(\rho)=v_0\rho(1-\rho)$ of a corresponding
  ASEP [$v_0=0.043$ from Eq.~\eqref{eq:v0-general}].}
\label{fig:current-cosine-potential}
\end{figure}

\begin{figure*}[t!]
\includegraphics[width=0.95\textwidth]{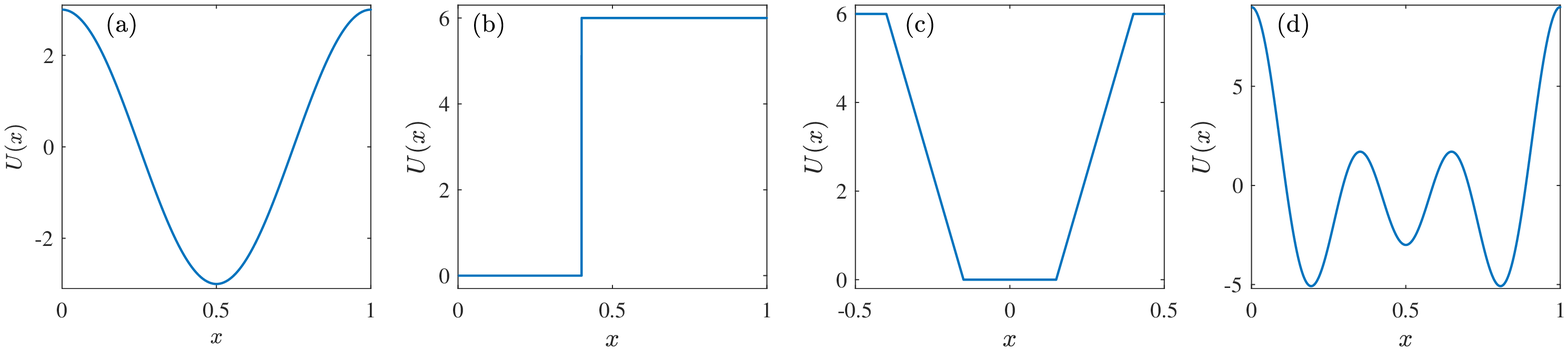}
\caption{The different periodic externals potentials
  investigated for comparing current-density relations: (a) cosine,
  (b) Kronig-Penney [Eq.~\eqref{eq:kronig-penney}], (c) piece-wise linear [Eq.~\eqref{eq:piecewise-potential}], and
  (d) triple-well [Eq.~\eqref{eq:triple-well-potential}].}
\label{fig:external-potentials}
\end{figure*}

\section{Hardcore interacting particles in harmonic potential}
\label{sec:reference-BASEP}
The paradigmatic variant of the BASEP with hardcore interacting
particles diffusing in the external harmonic potential
\begin{equation}
U(x) = \frac{U_0}{2} \cos\(\frac{2\pi x}{\lambda}\)
\label{eq:cosine-potential}
\end{equation}
has been studied thorougly in our previous works \cite{Lips/etal:2018,
  Ryabov/etal:2019, Lips/etal:2019}.  Here, we review its basic
properties that shall serve as a ``reference case'' for the following
analysis.

In all illustrations, we fix units setting $\lambda=1$ (defines units
of length), $\lambda^2/D=1$ (time), $\kB T=1$ (energy); this implies
that $\mu=D/(\kB T)=1$ also. We assume $U_0 \gg 1$, which leads to a
hopping-like motion between potential wells that resembles the
dynamics on a lattice.

Four representative shapes of current-density relations are shown in
Fig.~\ref{fig:current-cosine-potential}. In the low-density limit, all
curves collapse to the linear behavior $j_0(\rho) = v_0 \rho$ with the
slope given by the velocity $v_0$ of a single (non-interacting) particle.
This is given by \cite{Ambegaokar/Halperin:1969}
\begin{eqnarray}
v_0 = \frac{D\lambda(1-e^{-\beta f \lambda})}
{\int\limits_0^\lambda \dd x\! \int\limits_x^{x+\lambda}\! \dd y\, 
\exp[\beta(U(y)\!-\!fy\!-\!U(x)\!+\!fx)]}\,,
\label{eq:v0-general}
\end{eqnarray}
where $\beta=1/(\kB T)$.
Beyond the small-$\rho$ region, the shapes change strongly with the
particle size $\sigma$. This complex behavior is caused by three
competing collective effects:

{\em (i) The barrier reduction effect} leads to a current increase
with $\rho$. It appears in multi-occupied wells, where particles are
pushing each other to regions of higher potential energy and thus
decrease an effective barrier for a transition to neighboring
wells. The effect is best visible for small $\sigma$ causing currents
to be larger than $j_0(\rho)=v_0\rho$ (solid black line in
Fig.~\ref{fig:current-cosine-potential}). Likewise, for small and
moderate $\sigma$, the strong current increase at larger $\rho$ is due
to the occurrence of double-occupied wells.

{\em (ii) The blocking effect} suppresses the current by reducing the
number of transitions between neighboring wells. It occurs for larger
particle sizes: an extended particle is more easily blocked by another
one occupying the neighboring well (compared to smaller $\sigma$). To
contrast with the most extreme case of blocking, the parabolic
current-density relations $j_{\rm\scriptscriptstyle
  ASEP}(\rho)=v_0\rho(1-\rho)$ of a corresponding ASEP is shown as the
dashed line in Fig.~\ref{fig:current-cosine-potential}.

{\em (iii) The exchange symmetry effect} causes a deformation of the
current-density relation towards the linear behavior $j_0(\rho) = v_0
\rho$ if the particle size is close to $\sigma=m$, $m=0,1,2,\ldots$,
i.e.\ a multiple integer of $\lambda$.  In the commensurate case
$\sigma=m$, the current of interacting particles becomes equal to that
of noninteracting ones.  This effect is a consequence of the general
relation
\begin{equation}
\jss (\rho, \sigma) = (1-m\rho)\, \jss \left( \frac{\rho}{1-m\rho}, \sigma-m\lambda \right)
\end{equation}
that maps the stationary current in a system with particles of
diameter $\sigma$ and density $\rho$ to that with particles of
diameter $\sigma - m\lambda$ and density $\rho / (1-m\rho)$, where
$m={\rm int}(\sigma/\lambda)$ is the integer part of $\sigma/\lambda$.

\section{Impact of external periodic potential}
\label{sec:impact-external-potential}
As discussed in the Introduction, we consider further external
potentials, namely the Kronig-Penney, a piece-wise linear, and a triple-well
potential.  These potentials are plotted in
Fig.~\ref{fig:external-potentials}(b)-(d) together with the cosine
potential of our reference system in
Fig.~\ref{fig:external-potentials}(a).

\begin{figure}[b!]
\includegraphics[width=0.45\textwidth]{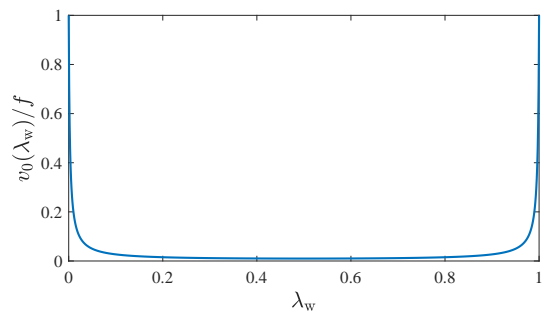}
\caption{Drift velocity of a single particle in dependence of $\lw$
  for the Kronig-Penney potential in Eq.~\eqref{eq:kronig-penney}; the
  velocity is normalized to $f$, i.e.\ its value in a flat (vanishing)
  external potential. The barrier height and drag force are $U_0=6$
  and $f=0.2$.}
\label{fig:velocity-kp}
\end{figure}

\subsection{Kronig-Penney potential}
\label{subsec:kronig-penney}
The Kronig-Penney potential has the form
\begin{equation}
U(x)=\left\{\begin{array}{ll}
0\,, & 0\le x<\lw\,,\\[1ex]
U_0\,, & \lw\le x< \lambda\,,
\end{array}\right.
\label{eq:kronig-penney}
\end{equation}
where $\lw$ is width of the rectangular well, and $\lb=\lambda-\lw$
the width of the rectangular barrier.  We are interested in the
current-density relation for different $\lw$ in the limit of large
$U_0\gg1$. Specifically, we take the same value $U_0=6$ as for the
reference BASEP with cosine potential discussed in
Sec.~\ref{sec:reference-BASEP}.  In particular, we aim to clarify,
whether a current enhancement over that of noninteracting particles
still occurs. As all particles dragged from one well to a neighboring
one have to surmount the same barrier height $U_0$ now, it is not
clear whether multiple occupation of wells lead to an effective
barrier reduction. The blocking and exchange symmetry effect are
expected to influence the current in an analogous manner.

\begin{figure*}[t!]
\includegraphics[width=0.95\textwidth]{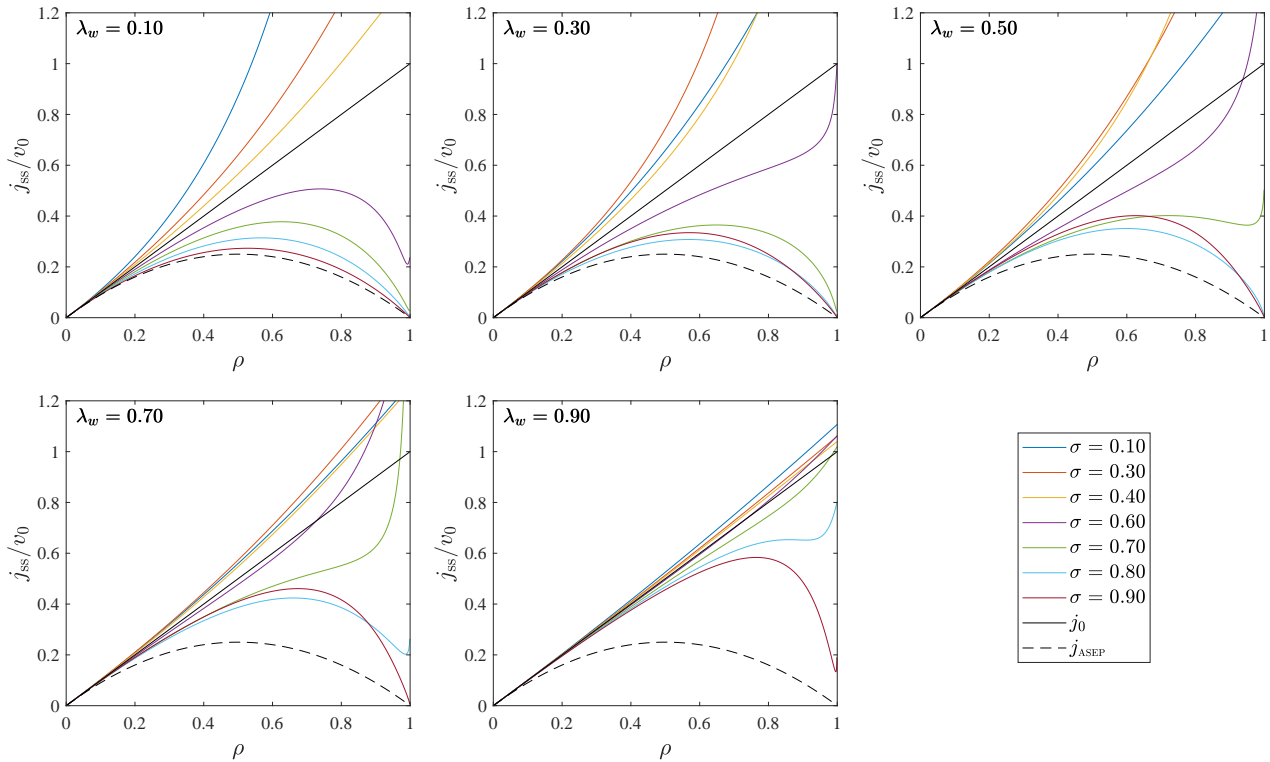}
\caption{Current $\jss$, normalized with respect to $v_0(\lw)$ (see
  Fig.~\ref{fig:velocity-kp}), for the Kronig-Penney potential in the
  small-driving approximation as a function of $\rho$ for different
  $\sigma$ and (a) $\lw=0.1$, (b) 0.3, (c) 0.5, (d) 0.7, and (e) 0.9.
  The barrier height is $U_0=6$.}
\label{fig:current-kp}
\end{figure*}

Inserting the Kronig-Penney potential in Eq.~\eqref{eq:v0-general}
yields 
\begin{subequations}
\begin{align}
v_0&(\lw)= \frac{A}{B}\\[0.5ex]
A&=\frac{D}{\lambda}(\beta f\lambda)^2(e^{\beta f \lambda}-1)\\[0.5ex]
B&=2\[e^{\beta f \lw}\!+\!e^{\beta f (\lambda\!-\!\lw)}\!-\!e^{\beta f \lambda}\!-\!1\]
\[1\!-\!\cosh(\beta U_0)\]\nonumber\\
&\hspace{1em}{}+\beta f\lambda(e^{\beta f \lambda}\!-\!1)
\label{eq:v0-kp}
\end{align}
\end{subequations}
for the single-particle velocity.  This result is plotted in
Fig.~\ref{fig:velocity-kp}. As expected, $v_0(\lw)$ approaches the
mean drift velocity $\mu f$ of a single particle in a flat potential
in the limits $\lw\to0$ (zero well width) and $\lw\to1$ (zero barrier
width).  With increasing width of the wells (or of the barriers),
$v_0(\lw)$ rapidly decreases.  Interestingly, Eq.~\eqref{eq:v0-kp}
implies the symmetry $v_0(\lw)=v_0(1-\lw)=v_0(\lb)$, that means the
single-particle velocity remains unaltered if the barriers and wells
are interchanged.

Current-density relations for hardcore interacting particles
calculated from the SDA (cf.\ Sec.~\ref{subsec:small-driving}) are
shown in Fig.~\ref{fig:current-kp} for five different value of
$\lw$. For each $\lw$, we plotted $\jss/v_0$ vs.\ $\rho$ for eight rod
lengths $\sigma$ analogous to our representation of current-density
curves in Fig.~\ref{fig:current-cosine-potential}. As can be seen from
the graphs, the shapes of the current-density relation are
qualitatively comparable to that in
Fig.~\ref{fig:current-cosine-potential} for all $\lw$, as well as
their overall change with the diameter $\sigma$. This means that the
interplay of the barrier reduction, blocking and exchange symmetry is
still present.  As expected, the overall strengths of the effects in
modifying the current of noninteracting particles becomes weaker with
decreasing $\lw$; for $\lw\to0$ the current $j(\rho,\lw)$ indeed
approaches $j_0(\rho,\lw\to0) =\rho v_0(\lw\to0)=\mu f\rho$.

\begin{figure}[t!]
\includegraphics[width=0.4\textwidth]{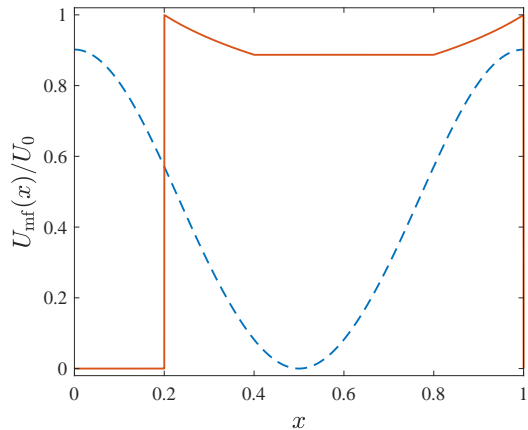}
\caption{Potential of mean force $U_{\rm mf}$ for the cosine potential
  (blue dashed line) and the Kronig-Penney potential with $\lw=0.2$
  (solid red line). Parameters are $\sigma=0.2$, $\rho=0.5$ and
  $U_0=6$ for both potentials.}
\label{fig:umf-kp}
\end{figure}

The barrier reduction, however, can no longer be associated with a
decrease of an effective barrier height, when two or more particles
occupy a potential well. For the Kronig-Potential in
Eq.~\eqref{eq:kronig-penney}, all particles in a multiple-occupied
well have zero energy and need to overcome $U_0$. Nevertheless, we can
attribute the enhancement of the current compared to that of
noninteracting particles with a barrier reduction. To see this, we
analyze the potential of mean force
\begin{equation}
\umf(x)=\kB T\ln\varrhoeq(x)+C=\umf(x+\lambda)
\end{equation}
for both the cosine and the Kronig-Penney potential, where the
constant $C$ is chosen to give a potential minimum equal to zero.  If
considering driven Brownian motion of noninteracting particles in the
potential $\umf(x)$, the current in the linear response limit would be
equal to the many-particle current in the SDA.  We therefore can
interpret $\umf$ as an effective barrier in the many-particle system.
In Fig.~\ref{fig:umf-kp}, we show $\umf/U_0$ for both the cosine and
the Kronig-Penney potential.  At the maximum of the cosine potential
at $x=0$, we find $\umf/U_0<1$, that means the barrier height is
reduced. In contrast, the barrier at the step of the Kronig-Penney
potential equals $U_0$ ($\umf(x=\lw)/U_0=1$ in
Fig.~\ref{fig:umf-kp}). However, a barrier reduction is now clearly
seen in the plateau part of the barrier in the range $\lw<x<1$. We
thus can distinguish between two types of barrier reduction, namely
the first type associated with a reduction of the barrier height and
the second type associated with a lowering of the barrier plateau.

Generally, a single-well periodic potential can be characterized
roughly by the widths of a valley and barrier part, and the flanks in
between these parts. A simple representation is given by the piecewise
linear potential
\begin{equation}
U(x)
=\left\{\begin{array}{cl}
0\,, & \displaystyle 0\le |x|\le\frac{\lw}{2}\,,\\[2ex]
\displaystyle\frac{U_0}{\lf}\left(|x|-\frac{\lw}{2}\right)\,, 
& \displaystyle\frac{\lw}{2}\le |x|\le\frac{\lw}{2}\!+\!\lf\,,\\[2ex]
U_0\,, & \displaystyle \frac{\lw}{2}\!+\!\lf\le |x|\le \lambda\,,
\end{array}\right.
\label{eq:piecewise-potential}
\end{equation}
shown in Fig.~\ref{fig:external-potentials}(c), where $\lw$, $\lf$ and $\lb=\lambda-\lw-2\lf$ are specifying the
widths of the valley, barrier and flanks. We performed additional
calculations in the SDA for this potential. For various fixed $\lw$
and $\lf$, we always found current-density relations with a behavior
similar to that found in Fig.~\ref{fig:current-cosine-potential} for
the BASEP with cosine potential. This model may thus be viewed as
representative for Brownian single-file transport through single-well
periodic potentials with barriers much larger than the thermal energy.

Current-density relations with a different characteristics can,
however, be obtained for multiple-well periodic potentials, as we
discuss next.

\begin{figure}[b!]
\includegraphics[width=0.45\textwidth]{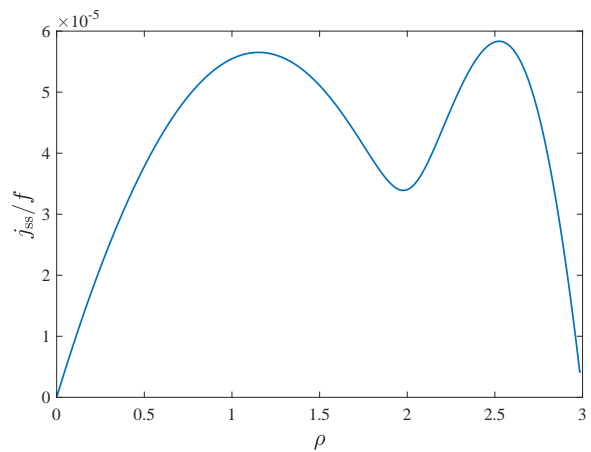}
\caption{Current $\jss$, normalized with respect to $f$, for the
  triple-well potential in the small-driving approximation as a
  function of $\rho$ for $\sigma=0.323$.  Densities at the local
  maxima are $\rho_{\rm max, 1}=1.14$ and $\rho_{\rm max, 2}=2.53$,
  and $\rho_{\rm min}=1.98$ at the local minimum.}
\label{fig:j-rho-triple-well}
\end{figure}

\subsection{Triple-well potential}
\label{subsec:triple-well}
The triple-well potential shown in
Fig.~\ref{fig:external-potentials}(d) is
\begin{equation}
U(x) = U_0\sum_{j=1}^3(j+1)\cos\(\frac{2\pi j x}{\lambda}\)\,,
\label{eq:triple-well-potential}
\end{equation}
where we choose $U_0=1$ here. For this potential, our calculations of
$\jss(\rho)$ based on the SDA show a very sensitive dependence on
$\sigma$. We concentrate here on one particle diameter $\sigma=0.323$,
where several local extrema occur, see
Fig.~\ref{fig:j-rho-triple-well}. In this figure currents are shown up
to a density (filling factor) $N/M=\rho=2.98$, corresponding to a
coverage $\rho\sigma\cong96\%$ of the system by the hard rods. If
$\rho$ approaches its maximal value $1/\sigma\cong 3.10$ corresponding
to a complete coverage, the numerical calculation of the equilibrium
density profile from Eq.~\eqref{eq:structure-equation} becomes
increasingly difficult. Hence, we refrained to show current data for
$\rho>2.98$ due to a lack of sufficient numerical accuracy when
calculating $\varrhoeq(x)$. It is important to state in this context
that the current for $\rho\to1/\sigma$ is expected to approach that of
noninteracting particles in a flat potential \cite{Lips/etal:2019},
i.e.\ it should hold $\jss(\rho,\sigma)\sim\mu f\rho$ for
$\rho\to1/\sigma$ [except for the singular point $\rho=1/\sigma$, where
  $\jss(1/\sigma,\sigma)=0$]. This means that the current in
Fig.~\ref{fig:j-rho-triple-well} must steeply rise for
$\rho\to1/\sigma$, i.e.\ there must appear a further local minimum for
$\rho>2.98$. These arguments apply also to the currents shown in
Figs.~\ref{fig:current-cosine-potential} and \ref{fig:current-kp}.

To explain the occurrence of the local extrema in
Fig.~\ref{fig:j-rho-triple-well}, we resort to
Eq.~\eqref{eq:jst_linear} with $\chi=0$, i.e.\ the SDA. This equation
can be interpreted by considering $\mu\varrhoeq(x)\dd
x/\lambda$ to be the ``local conductivity'' of a line segment $\dd
x$. A serial connection of these segments implies that the ``total
conductivity'' is given by the inverse of the sum of the inverse local
conductivities, corresponding to a summation of the respective ``local
resistivities''. A stronger localization of $\rho_\textrm{eq}(x)$
around the minima of the potential leads to a smaller conductivity and
hence a smaller current $\jss(\rho)$, while less localized density
profiles lead to larger $\jss(\rho)$.

Using this picture, the occurrence of the first maximum in
$\jss(\rho)$ can be traced back to an increasing occurrence of double
occupied wells for $\rho\gtrsim1$.  In double-occupied wells, particle
motion is more restricted, leading to a stronger particle localization
at the two deeper minima at about $x\simeq0.2$ and $x\simeq0.8$, see
Fig.~\ref{fig:external-potentials}(d).  Accordingly, $\jss(\rho)$
starts to decrease with $\rho$ for $\rho\gtrsim 1$ [$\rho_{\rm max,
    1}\cong1.14$ in Fig.~\ref{fig:j-rho-triple-well}]. The
decrease of $\jss(\rho)$ continues up to a filling factor of about two
[$\rho_{\rm min}\cong1.98$ in Fig.~\ref{fig:j-rho-triple-well}],
above which more than two particles occupy a well on average.  With a
significant appearance of triple-occupied wells goes along first a
stronger spreading of the density, as the minimum of the potential at
$x=1/2$ becomes occupied in wells containing three particles.  The
spreading of the density causes $\jss(\rho)$ to increase for
$\rho\gtrsim2$. A counteracting effect, however, is a strong particle
localization at all potential minima in neighboring triple-occupied
wells, where the hardcore constraints force the particles to become
strongly localized around the potential minima.  For $\rho\gtrsim2.5$
[$\rho_{\rm max, 2}\cong2.53$ in Fig.~\ref{fig:j-rho-triple-well}], every second well is
occupied by three particles on average which lets $\jss(\rho)$ to
decrease again with further increasing $\rho$.

The more complex current-density relation in
Fig.~\ref{fig:j-rho-triple-well} leads to a richer variety of NESS
phases in an open systems compared to the reference BASEP, which can
exhibit up to five different phases \cite{Lips/etal:2018,
  Lips/etal:2019}.  To identify all possible NESS phases, we consider
the particle exchange with two reservoirs L and R at the left and
right end of an open system to be controlled by two parameters $\rhoL$
and $\rhoR$. As discussed in connection with
Eq.~\eqref{eq:extremal-current-principles} in the Introduction, these
control parameters can be considered as effective densities, or they
can be associated with true reservoir densities for specific
bulk-adapted couplings of the system to the reservoirs
\cite{Dierl/etal:2013, Dierl/etal:2014}.

\begin{figure}[t!]
\includegraphics[width=0.45\textwidth]{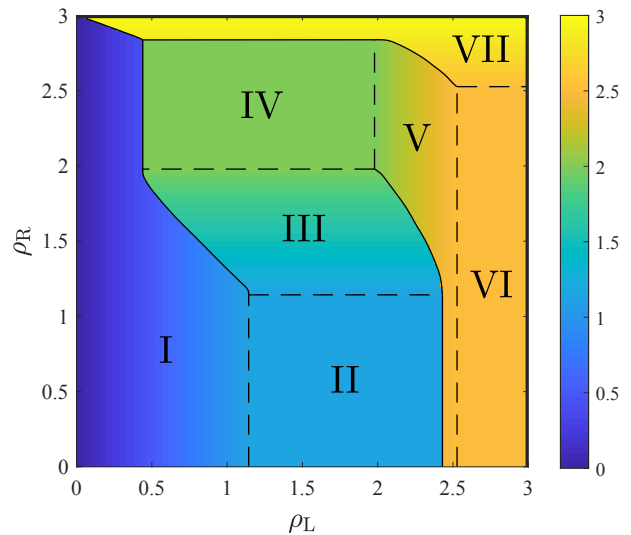}
\caption{Phase diagram of NESS for the triple-well potential obtained
  by applying the extremal current principles to the current-density
  relation in Fig.~\ref{fig:j-rho-triple-well}. The color bar encodes
  the values of the bulk density $\rhoB$ in an open system coupled to
  particle reservoirs. Phases I and V are left-boundary matching
  phases with $\rhoB=\rhoL$, phases III and VII are right-boundary
  matching phases with $\rhoB=\rhoR$, phases II and VI are maximal
  current phases with $\rhoB=\rho_{\rm max,1}$ and $\rhoB=\rho_{\rm
    max,2}$ respectively, and phase IV is a minimal current phase with
  $\rhoB=\rho_{\rm min}$. Solid (dashed) lines mark first (second)
  order phase transitions. The dark black bars at the boundaries mark
  the two stripes $2.98<\rho_{\rms L,R}\le1/\sigma\cong3.1$, where
  additional phases appear (see the discussion in
  Sec.~\ref{subsec:triple-well}).}
\label{fig:phase-diagram-triple-well}
\end{figure}

Applying Eq.~\eqref{eq:extremal-current-principles} with
$\rho_-=\rhoL$ and $\rho_+=\rhoR$ to the current-density relation in
Fig.~\ref{fig:j-rho-triple-well} results in the diagram with seven
different NESS phases I-VII shown in
Fig.~\ref{fig:phase-diagram-triple-well}.  The color coding shows the
value of the bulk density $\rhoB$, i.e.\ the order parameter of the
phase transitions.  Solid lines mark first order and dashed lines
second order phase transitions, which is reflected in the smooth
(continuous) or sudden (jump-like) changes of the color.  The seven
phases can be classified in two categories: boundary-matching phases,
where $\rhoB$ is equal to either $\rhoL$ or $\rhoR$, and extremal
current phases, where $\rhoB$ is equal to one of the densities, where
$\jss(\rho)$ has a local extremum in Fig.~\ref{fig:j-rho-triple-well}.
Specifically, the phases I and V are left-boundary matching phases
with $\rhoB=\rhoL$, the phases III and VII are right-boundary matching
phases with $\rhoB=\rhoR$, phase II is a maximal current phase with
$\rhoB=\rho_{\rm max,1}$, phase VI a maximal current phase with
$\rhoB=\rho_{\rm max,2}$, and phase IV a minimal current phase with
$\rhoB=\rho_{\rm min}$. If one takes into account the existence of the
further minimum for $\rho>2.98$ in the current-density relation (see
discussion above), then even more phases are possible.  These
additional phases, however, must appear in the two stripes
$2.98<\rho_{\rms L,R}\le1/\sigma\cong3.1$, i.e.\ in a very narrow
range of one of the two control parameters (marked in black in
Fig.~\ref{fig:phase-diagram-triple-well}).

\section{Impact of interactions other than hardcore exclusions}
\label{sec:impact-interactions}
In this section, we investigate the impact of other particle
interactions beyond hardcore exclusion for the cosine external
potential in Eq.~\eqref{eq:cosine-potential}. This is done in two
different settings. First, we investigate the repulsive Yukawa
potential
\begin{equation}
\uintY(r) = \AY\,\frac{e^{-r/\xi}}{r/\xi} \, ,
\label{eq:yukawa-potential}
\end{equation}
between particles at distance $r$ for a fixed small amplitude $\AY=1$
and different decay length $\xi$.  Secondly, we combine this Yukawa
interaction with hardcore interactions. For obtaining current-density
relations, we here employ Brownian dynamics simulations. This is
because an exact density functional is not available for the Yukawa
interaction and the SDA with a precise determination of $\varrhoeq(x)$
cannot be applied. For performing the Brownian dynamics simulations,
we used a standard Euler integration scheme of the Langevin equations
\eqref{eq:langevin} with a time step $\Delta t = 10^{-4}$.  To deal
with the hardcore interactions, the algorithm developed in
Ref.~\onlinecite{Scala:2012} was applied.

\begin{figure}[t!]
\includegraphics[width=0.45\textwidth]{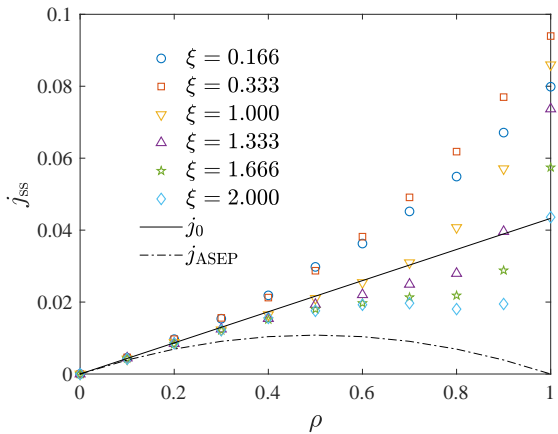}
\caption{Simulated current-density relations for the external cosine
  potential [Eq.~\eqref{eq:cosine-potential}], and particle
  interactions given by the Yukawa potential in
  Eq.~\eqref{eq:yukawa-potential}.  The amplitude of the Yukawa
  potential is $\AY=1$ and $\xi$ values specify different decay
  lengths.  The other simulation parameters are $U_0=6$ and $f=2$. The
  solid line marks the current $j_0(\rho)=v_0\rho$ for noninteracting
  particles and the dashed line the current-density relation $j_{\rms
    ASEP}(\rho)=v_0\rho(1-\rho)$ of a corresponding ASEP (same curves
  as in Fig.~\ref{fig:current-cosine-potential}).}
	\label{fig:yukawa-current-density}
\end{figure}

Current-density relations for the Yukawa potential without additional
hardcore interactions are shown in
Fig.~\ref{fig:yukawa-current-density} for six different values of
$\xi$. These may be viewed to resemble effective particle diameters
$\sigma$ of hardcore interacting systems.  For small $\xi$,
i.e.\ $\xi=0.166$ and $0.333$ in
Fig.~\ref{fig:yukawa-current-density}, $\jss$ shows an enhancement
over that of noninteracting particles (solid black line) due to a
prevailing barrier reduction effect similar to that in the reference
BASEP for small $\sigma$. When enlarging $\xi$, the current is reduced
for small $\rho$ compared to that of noninteracting particles, while
it rises strongly for large $\rho$. Again this behavior is analogous
to that in the reference BASEP for increasing $\sigma$. Because the
(effective) blocking effect is not so strong for $\AY=1$, the
current-density curves do not approach the limiting
$j_{\scriptscriptstyle \rm ASEP}(\rho)$ as closely as for hardcore
interactions. Nevertheless, one can say that the change of the
current-density relation with varying $\xi$ is reflecting the
interplay of a barrier reduction and blocking effect as in the
reference BASEP.

\begin{figure}[t!]
\includegraphics[width=0.45\textwidth]{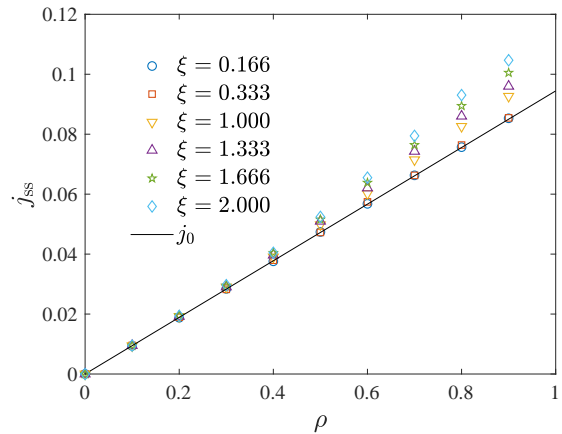}
\caption{Simulated current-density relations for the cosine external
  potential and Yukawa interactions as in
  Fig.~\ref{fig:yukawa-current-density}, and additional hardcore
  interactions for commensurate particle size $\sigma=1$. Parameters
  for the drag force as well as for the amplitudes of the cosine and
  Yukawa potential are chosen as in
  Fig.~\ref{fig:yukawa-current-density}. The solid line marks the
  current $j_0(\rho)=v_0\rho$ for noninteracting particles (same line
  as in Figs.~\ref{fig:current-cosine-potential} and
  \ref{fig:yukawa-current-density}.)}
	\label{fig:yukawa-hardcore-current-density}
\end{figure}

However, one cannot find certain $\xi$ values, where the
current-density relation equals that of noninteracting particles for
all $\rho$.  The peculiar exchange symmetry effect in the BASEP for
commensurate $\sigma=m$, $m=0,1,2\ldots$, is caused by the invariance
of the stochastic particle dynamics against a specific coordinate
transformation containing $\sigma$ \cite{Lips/etal:2018,
  Lips/etal:2019}. Such coordinate transformation does not exist for
the Yukawa potential.

For the Yukawa potential with additional hardcore interactions, we
found changes of current-density curves caused by the barrier
reduction and blocking effect as discussed above. But it is
interesting to analyze now, whether the relation
$\jss(\rho)=j_0(\rho)=v_0\rho$ for commensurate $\sigma=m$ and $\xi=0$
is approximately reflected in current-density relations for $\xi>0$,
where the exchange symmetry effect is no longer strictly valid.  One
may expect that the hardcore interacting system should be only weakly
perturbed by the Yukawa potential if $\AY$ is of the order of the
thermal energy and $\xi$ not too large compared to $\sigma$.  This is
indeed confirmed by simulation results for $\sigma=1$ shown in
Fig.~\ref{fig:yukawa-hardcore-current-density}. The data points for
$\xi = 0.166$ and $\xi=0.333$ lie almost directly on the curve
$j_0(\rho)$ up to the highest simulated density $\rho=0.9$. With
increasing $\xi$, deviations from the linear behavior are seen, which
become the more pronounced the larger $\rho$. But even for $\xi=2$,
$\jss(\rho)$ follows $j_0(\rho)$ closely up to $\rho=0.4$.  We thus
conclude that slight deviations from a perfect hardcore interaction,
as they are always present in experiments, still allow an
identification of the exchange symmetry effect.

\section{Summary and Conclusions}
\label{sec:conclusions}
To analyze how generic our previous findings are for the
nonequilibrium physics of the BASEP in a sinusoidal potential, we have
studied the driven Brownian motion of hardcore interacting particles
for other external periodic potentials.  Our calculations were carried
out based on a small-driving approximation, which refers to the linear
response under neglect of a period-averaged mean interaction force.
If the external periodic potential exhibits a singe-well structure
between barriers, i.e.\ if there is just one local minimum per period,
our results provide evidence that the various characteristic shapes of
bulk current-density relations $\jss(\rho)$ for different particle
sizes $\sigma$ are always occurring. There are differences in the
exact functional form and at which $\sigma$ the shape type is
changing.  For all single-well periodic potentials it is the interplay
of a barrier reduction, blocking and exchange symmetry effect that
causes a particular shape type to appear. Even for a Kronig-Penney
potential with alternating rectangular well and barrier parts, where
the barrier reduction effect is not so obvious, we showed that an
enhancement of the current over that of noninteracting particles
occurs. For that potential this enhancement can be attributed to an
effective reduction of the barrier plateau parts. The generic behavior
of the bulk current-density relations implies that for single-well
periodic potentials up to five different NESS phases appear in open
BASEP systems coupled to particle reservoirs. This can be concluded by
applying the extremal current principles \cite{Kolomeisky/etal:1998,
  Popkov/Schuetz:1999}.

More complex shape types of $\jss(\rho)$ can occur in multiple-well
periodic potentials. This was demonstrated for a particular
triple-well potential, where our calculations yielded a
current-density relation with two local maxima for a certain particle
size. In that case the extremal current principles predict more than
five different NESS in an open system.  When neglecting a very narrow
range of effective reservoir densities, which would be very difficult
to realize by specific system-reservoir couplings in simulations or
experiments, up to seven different NESS phases are possible. We point
out that these results were obtained here for demonstration
purposes. Systematic investigations of multiple-well external
potentials should be performed in the future with a goal to reach a
general classification similar as for the BASEP for single-well
periodic potentials.

Current-density relations with several local maxima are particularly
interesting in the case of ``degenerate maxima''. i.e.\ when the
current at the maxima has the same value. In such situations,
coexisting NESS phases of maximal current can occur in a whole
connected region of the space spanned by the parameters controlling
the coupling to the environment \cite{Maass/etal:2018}. Such states of
coexisting extremal current phases have not yet been studied in detail
in the literature. Preliminary results for driven lattice gases
indicate that fluctuations of interfaces separating extremal current
phases exhibit an anomalous scaling with time and system length
\footnote{D.~Locher, 
bachelor thesis (in German), Osnabr{\"u}ck University (2018); M.~Bosi,
D.~Locher, and P. ~Maass, to be published.}.  This is in contrast to
the already well-studied interface fluctuations between the low- and
high-density phases in the ASEP, which at long times show a simple
random-walk behavior.

We furthermore performed Brownian dynamics simulations of driven
single-file diffusion through a cosine potential for a repulsive
particle interaction other than hardcore exclusion. Specifically, we
chose a Yukawa interaction with a small interaction amplitude equal to
the thermal energy and studied the behavior for different decay
lengths $\xi$. Current-density relations for this system showed
similar shapes as for the BASEP except for the effects implied by the
exchange symmetry effect, which is absent for other interactions than
hardcore.  The change of shapes is solely determined by the interplay
of a barrier reduction and effective blocking effect.  If the hardcore
interaction and the weak Yukawa interaction are combined, the
consequences of the exchange symmetry effect can be still seen for
particle sizes commensurate with the wavelength of the cosine
potential. The current $\jss(\rho)$ follows closely that of
noninteracting particles up to high densities even for large
$\xi$. This means that deviations from a perfect hardcore interaction
in experiments should still allow one to verify the exchange symmetry.

\vspace{2ex}
\acknowledgments
\vspace{-2ex} 
Financial support by the Czech Science Foundation (Project
No.\ 20-24748J) and the Deutsche Forschungsgemeinschaft (Project
No.\ 397157593) is gratefully acknowledged. We sincerely thank the
members of the DFG Research Unit FOR 2692 for fruitful discussions.

\appendix*
\section{Example for connection of ASEP to quantum spin chain}
Let $p(n,t)$ denote the probability of configurations $n=\{n_i,
i=1,\ldots,L\}$ of occupation numbers $n_i\in\{0,1\}$ in a
single-species fermionic lattice gas at time $t$.  Its time evolution
is described by the master equation
\begin{align}
\frac{\dd p(n,t)}{\dd t}&=\sum_{n'} [w_{nn'}p(n',t)-w_{n'n}p(n,t)]\nonumber\\
&=\sum_{n'} H_{nn'}p(n',t)
\label{eq:master}
\end{align}
where $w _{nn'}$ is the transition rate from configuration $n'$ to $n$
(for $n'\ne n$, otherwise $w_{nn}=0$), and
$H_{nn'}=w_{nn'}-\delta_{nn'}\sum_{n''}w_{n''n}$.  This master
equation corresponds to the occupation number representation of a
Schr\"odinger equation
\begin{equation}
\frac{\dd\ket{p}}{\dd t}=H\ket{p}
\label{eq:schroedinger}
\end{equation}
in imaginary time \cite{Gwa/Spohn:1992, Sandow/Trimper:1993}.

For the ASEP with periodic boundary conditions ($n_{L+1}=n_1$,
$n_0=n_L$), the transitions rates $w_{nn'}$ can be written as
\begin{align}
w_{nn'}&=\sum_{j=1}^L \delta_{n'n^{(j)}}[\Gamma_-n_{j+1}'(1\!-\!n_j')\!+\!\Gamma_+n_j'(1\!-\!n_{j+1}')]\nonumber\\
&=\sum_{j=1}^L\delta_{n'n^{(j)}}[\Gamma_-(1\!-\!n_{j+1})n_j\!+\!\Gamma_+(1\!-\!n_j)n_{j+1}]\,,
\label{eq:wnn}
\end{align}
where $n^{(j)}$ denotes the configuration $n$ with the occupation
numbers at sites $j$ and $(j\!+\!1)$ interchanged, i.e.,
$n^{(j)}_k=n_k$ for $k\ne j,(j+1)$, $n^{(j)}_j=n_{j+1}$, and
$n^{(j)}_{j+1}=n_j$.  Accordingly, the matrix elements
$H_{nn'}=\bra{n}H\ket{n'}=w_{nn'}-\delta_{nn'}\sum_{n''}w_{n''n}$ are
\begin{align}
H_{nn'}=&\sum_{j=1}^L\bigl\{\delta_{n'n^{(j)}}[\Gamma_-n_j(1\!-\!n_{j+1})\!+\!\Gamma_+(1-n_j)n_{j+1}]\nonumber\\
&{}-\delta_{nn'}[\Gamma_-(1\!-\!n_j)n_{j+1}\!+\!\Gamma_+n_j(1\!-\!n_{j+1})]\bigr\}\,.
\label{eq:Hnn}
\end{align}
Because $\bra{n}c_j^\dagger
c_{j+1}\ket{n'}=\delta_{n'n^{(j)}}n_j(1\!-\!n_{j+1})$ for creation and
annihilation operators $c_j^\dagger$ and $c_j$ of a particle at site
$j$, the matrix elements in Eq.~\eqref{eq:Hnn} are equal to that of
the Hamiltonian
\begin{align}
H=&\sum_{j=1}^L \bigl\{\Gamma_-[c_j^\dagger c_{j+1}-n_{j+1}(1\!-\!n_j)]\nonumber\\
&\hspace{2em}{}+\Gamma_+[c_{j+1}^\dagger c_j-n_j(1\!-\!n_{j+1})]\bigr\}
\label{eq:H-1}
\end{align}
of spinless fermions, which for $\Gamma_+\ne\Gamma_-$ is
non-Hermitian. In a representation by Pauli matrices, one can write
$c_j^\dagger=\sigma_j^+/2=(\sigma_j^x\!+\!i\sigma_j^y)/2$,
$c_j=\sigma_j^-/2=(\sigma_j^x\!-\!i\sigma_j^y)/2$,
$n_j=(1+\sigma_j^z)/2$, giving
\begin{align}
H=\frac{1}{4}&\sum_{j=1}^L \bigl[\Gamma_-\sigma_j^+ \sigma_{j+1}^-+\Gamma_+\sigma_{j+1}^+\sigma_j^-\nonumber\\
&\hspace{2em}{}+(\Gamma_-\!+\!\Gamma_+)(\sigma_j^z\sigma_{j+1}^z-1)\bigr]\,.
\label{eq:H-2}
\end{align}
The periodic boundary conditions imply $\sigma_{L+1}^\pm=\sigma_1^\pm$
and $\sigma_{L+1}^z=\sigma_1^z$.

A transformed $H'=VHV^{-1}$ with (non-singular) operator $V$ has the
same spectrum as $H$, where eigenstates $\ket{\varphi}$ and
$\ket{\varphi'}$ of $H$ and $H'$ to the same eigenvalue are related by
$\ket{\varphi'}\!=\!V\ket{\varphi}$.  Such transformation can be used
to symmetrize the non-Hermitian part $(\Gamma_-\sigma_j^+
\sigma_{j+1}^-+\Gamma_+\sigma_{j+1}^+\sigma_j^-)$ in
Eq.~\eqref{eq:H-2} by choosing
$V\!=\!\exp(\alpha\sum_{j=1}^Lj\sigma_j^z)$ with some constant
$\alpha$, because $V\sigma_j^{\pm}V^{-1}\!=\!e^{\pm2\alpha
  j}\sigma_j^\pm$ \cite{Henkel/Schuetz:1994}.  With $V\sigma_j^+
\sigma_{j+1}^-V^{-1}\!=\!V\sigma_j^+
VV^{-1}\sigma_{j+1}^-V^{-1}\!=\!e^{-2\alpha}\sigma_j^+ \sigma_{j+1}^-$
and
$V\sigma_{j+1}^+\sigma_j^-V^{-1}\!=\!e^{2\alpha}\sigma_{j+1}^+\sigma_j^-$,
the symmetrization is achieved by requiring
$\Gamma_-e^{-2\alpha}=\Gamma_+e^{2\alpha}$, i.e.\ by setting
$e^{\alpha}\!=\!(\Gamma_-/\Gamma_+)^{1/4}$. The transformed
Hamiltonian is that of a quantum $XXZ$ chain,
\begin{align}
H'=&\frac{\sqrt{\Gamma_-\Gamma_+}}{2}\sum_{j=1}^L \bigl(\sigma_j^x\sigma_{j+1}^x+\sigma_j^y\sigma_{j+1}^y+\Delta\,\sigma_j^z\sigma_{j+1}^z-\Delta\bigr)
\label{eq:Hprime}
\end{align}
with $\Delta=(\Gamma_-\!+\!\Gamma_+)/(2\sqrt{\Gamma_-\Gamma_+})$, but
now non-Hermitian boundary conditions $V\sigma_{L+1}^\pm
V^{-1}=V\sigma_1^\pm V^{-1}$, i.e.\ $\sigma_{L+1}^\pm=e^{\mp2\alpha
  L}\sigma_1^\pm=(\Gamma_-/\Gamma_+)^{\mp L/2}\sigma_1^\pm$ and
$\sigma_{L+1}^z=\sigma_1^z$.



%

\end{document}